\documentclass[twocolumn,showpacs,preprintnumbers,amsmath,amssymb,prl,tightenlines,final]{revtex4}
\usepackage{hyperref,graphics} 

\usepackage{graphicx}
\usepackage{dcolumn}
\usepackage{bm}
\usepackage{epsfig}
\usepackage{longtable}

\begin{document}

\title{Self-interaction errors in density functional calculations of
electronic transport}
\date{\today} 
\author{C. Toher$^1$, A. Filippetti$^2$, and S. Sanvito$^1$}
\affiliation{1) Department of Physics, Trinity College, Dublin 2, Ireland}
\affiliation{2) Sardinian Lab for Computational Material Science,
Physics Department, University of Cagliari,
I-09042 Monserrato (Ca), Italy}
\author{Kieron Burke}
\affiliation{Department of Chemistry and Chemical Biology, Rutgers University, 610 Taylor Rd., 
Piscataway, NJ 08854, USA}

\begin{abstract}
All density functional calculations of single-molecule transport to date
have used continuous exchange-correlation approximations. The lack of
derivative discontinuity in such calculations leads to the erroneous
prediction of metallic transport for insulating molecules. A simple and
computationally undemanding atomic self-interaction correction greatly
improves the agreement with experiment for the prototype
Au/dithiolated-benzene/Au junction.
\end{abstract}


\maketitle
Molecular devices are becoming increasingly important in a wide 
spectrum of applications. These range from the building blocks of revolutionary 
computer architectures \cite{gates1}, to disposable electronics, to diagnostic tools for genetically 
driven medicine, to multifunctional sensors \cite{chem}.

Therefore interest is growing in the development of computational tools
capable of predicting the $I$-$V$ characteristics of devices
comprising only a handful of atoms. In general
these are based on Landauer scattering theory, typically
in the non-equilibrium Green's function
(NEGF) formalism \cite{negf}, combined with an electronic structure method, usually density 
functional theory (DFT) \cite{dft,ks}.  Such schemes are physically appealing and yield
useful results \cite{smeagol}, even if they are incomplete \cite{EWK04,kieron,SZVD05},
and are computationally simpler than many-body methods \cite{MB}. 
The fundamental requirements for an electronic structure theory applied to the
problem of transport through single molecules
are: 1) to be accurate when the molecule has a fractional 
number of electrons, 2) to describe properly both the electron affinity ($A$) and ionization 
potential ($I$) of the isolated
molecule, 3) to be cast in a single particle form. The first two 
conditions are necessary for a correct description of the transport, while the third produces
computationally efficient algorithms.
Most importantly, as we show here, we need an accurate description of the HOMO state as 
a function of its occupation.

The exact Kohn-Sham (KS) potential of a $N$-electron system always satisfies the condition
$\epsilon_\mathrm{HOMO}^\mathrm{KS}= -I_N$, i.e., the highest occupied KS molecular
orbital energy is the negative of the $N$-electron ionization potential. Let $N+n$ be the
number of electrons localized on a molecule weakly coupled to a reservoir, where $N$
is an integer, but $n$ is continuous.  For $-1 <
n \leq 0$, $\epsilon_\mathrm{HOMO}^\mathrm{KS}= -I_N$, 
but for $0 <
n \leq 1$, $\epsilon_\mathrm{HOMO}^\mathrm{KS}= -I_{N+1}$.  To achieve this,
the KS potential jumps by a step of $I_N-I_{N+1} = I_N - A_N$, where $A_N$ is the
electron affinity.  This is the infamous derivative discontinuity
of DFT \cite{janak,PPLB}, which is missing  in ordinary continuous functionals such as
the local density (LDA) or the generalized gradient approximation (GGA) \cite{PL2}.
Smooth exchange-correlation functionals 
continuously connect the orbital levels for 
different integer occupations, leading to qualitative errors such as the 
erroneous prediction of the dissociation of heteronuclear molecules
into fractionally charged ions \cite{Pb85}. 
We now show the errors that this gives rise to in a typical transport calculation.

\begin{figure}[ht]
\begin{center}
\includegraphics[width=4.0cm,clip=true]{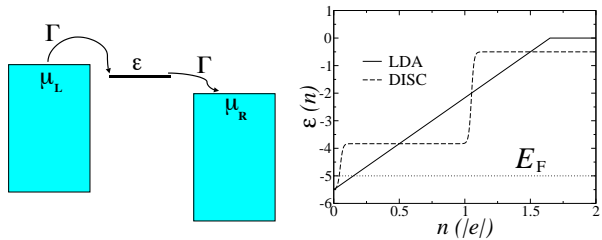}
\includegraphics[width=4.0cm,clip=true]{./Fig1b}
\end{center}
\caption{\small{(a) Schematic two terminal device. Two leads are kept at the chemical potentials
$\mu_\mathrm{L}$ and $\mu_\mathrm{R}$ and the transport is through a single energy level $\epsilon$.
The hopping energy between the leads and the energy level is $\Gamma$. (b) Dependence
of $\epsilon$ on its own occupation $n$. The straight line corresponds to
a typical LDA dependence and the step-like line to the DISC. Notice that the LDA line becomes flat 
at $n\sim$1.5 ($\epsilon(n)$=0) since here the eigenstate is unbound. The dotted horizontal line
denotes the position of the leads Fermi level $E_\mathrm{F}$.}}
\label{Fig1}
\end{figure}

We model a two terminal molecular device as two featureless leads (constant density of states) 
kept at different chemical potentials $\mu_\mathrm{L}$ and $\mu_\mathrm{R}$ and
coupled through a single energy level $\epsilon$ (Fig. \ref{Fig1}(a)). 
The density of states (DOS) associated with $\epsilon$ is a Lorentzian,
$D(E)=\frac{1}{2\pi}\frac{\Gamma}{(E-\epsilon)^2+(\Gamma/2)^2}$,
with broadening $\Gamma$ arising from the hopping to the leads.
The energy level occupation $n$ and the steady state current 
$I$ can be obtained by balancing the in-going and out-going currents to and from the energy 
level \cite{negf}. At steady state, $n$ is just proportional to the Fermi 
distributions $f(\epsilon,T)$ of the leads:
$n=\int_{-\infty}^{\infty}\mathrm{d}E\:D(E)[f(E-\mu_\mathrm{L},T)+f(E-\mu_\mathrm{R},T)]$, 
while the current is given by
\begin{equation}
I=\frac{e}{\hbar}\ \Gamma\int_{-\infty}^{\infty}\mathrm{d}E\:D(E)[f(E-\mu_\mathrm{L},T)-
f(E-\mu_\mathrm{R},T)]\;.
\end{equation}
The dynamics of this model are
rather simple. Consider the weak coupling limit ($\Gamma\ll A_N$), where simply 
$D(E)\sim\delta(E-\epsilon)$. Both occupation and current
are solely determined by the position of the energy level with respect to the chemical potential of the leads.
If $\epsilon$ is larger than both $\mu_\mathrm{L}$ and $\mu_\mathrm{R}$, then $n\approx0$
and no current flows. In contrast,
if the energy level is below the chemical potentials of both leads,
then $n\approx2$, but the current is still zero.
Finally, if $\mu_\mathrm{R}<\epsilon<\mu_\mathrm{L}$ the occupation is $0<n<2$
and current flows. Considering now
that the chemical potential in the leads is simply $\mu_\mathrm{L/R}=E_\mathrm{F}\pm eV/2$, 
where $E_\mathrm{F}$ is the Fermi level of both leads
and $V$ is the applied bias, this simple model predicts a conductance gap in the $I$-$V$ curve 
for $-2|E_\mathrm{F}-\epsilon|<eV<2|E_\mathrm{F}-\epsilon|$.

However, because this is an effective one-body representation of an interacting system,
in general the position of the energy level depends on its own occupation,
$\epsilon=\epsilon(n)$. Let us now solve this problem within KS
DFT \cite{ks}. For definiteness, assume that $\epsilon$ is the LUMO ($n$=0) of 
a certain molecule, which contains $N$ electrons in the neutral state. In the exact KS theory, 
when this molecule is weakly coupled to a reservoir, $\epsilon$ will be a 
discontinuous function of $n$ \cite{Pb85}. We parameterized this dependence 
with the step-like curve of figure \ref{Fig1}(b), 
and we call it ``DISC'' (discontinuous occupation).
For $0 < n \leq 1$, $\epsilon(n)=-A_N$, where $A_N$ is the electron affinity
of the isolated molecule, while $\epsilon(n)$ jumps rapidly to
its next plateau ($-A_{N+1}$) just above 1.
In contrast, the LDA energy level position $\epsilon$ varies approximately
linearly with $n$ ($\epsilon=U\; n$), reflecting the fact that the LDA total energy varies
approximately quadratically around the neutral configuration \cite{Alessio}. 

\begin{figure}[ht]
\begin{center}
\includegraphics[width=7.0cm,clip=true]{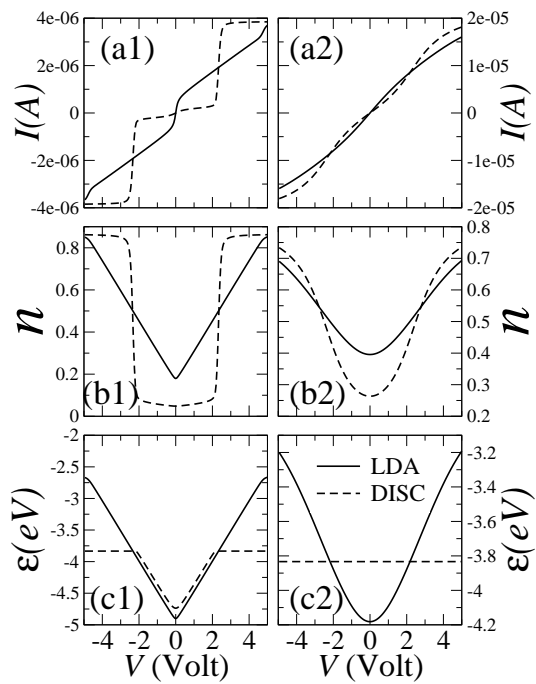}
\end{center}
\caption{\small{(a) Current $I$, (b) occupation $n$ and
(c) position of the energy level $\epsilon$ as a function of bias $V$. 
The parameters used here are $\epsilon(0)$=-5.5~eV, $U$=5~eV, 
$E_\mathrm{F}$=-5.0~eV and $T$=300$^o$K. The curves on the left-hand side are obtained in 
the weak coupling limit ($\Gamma$=0.2~eV) and those on the right-hand side in the 
strong coupling limit ($\Gamma$=1.2~eV).}}
\label{Fig2}
\end{figure}

The different $I$-$V$ characteristics that one obtains by using either the LDA or DISC
parameterizations are presented in Fig. \ref{Fig2} along with the level occupation and 
its position as a function of bias. These have been obtained by iterating 
self-consistently $n(\epsilon)$ with $\epsilon(n)$, where we assume $\epsilon(0)$ 
just below $E_\mathrm{F}$ ($|\epsilon(0)-E_\mathrm{F}|=0.5$~eV). Consider first 
the weak coupling limit. In both LDA and DISC, the energy level pins $E_\mathrm{F}$ 
at zero bias. As the bias is further increased, more charge fills the energy level, 
which keeps rising up. Figure \ref{Fig2}(c1) shows that this rise is found both 
in LDA and DISC and is approximately linear with the bias. Importantly, as 
soon as $\epsilon$ shifts above the chemical potential of the right-hand side 
contact, then $f(\epsilon-\mu_\mathrm{R})\approx0$, and the current will be 
simply proportional to the level occupation ($I\approx\Gamma f(\epsilon-\mu_\mathrm{L})$). 

Clearly LDA and DISC behave in a qualitatively different way. In fact, a 
LDA-type potential leads to a linear dependence of the occupation on bias
(see Fig. \ref{Fig2}(b1)), and consequently to a metallic conductance.
In contrast, in DISC, the energy level shifts upwards without substantial charging. 
The result is that the occupation jumps almost discontinusly from $n=0$ to $n=1$
when the bias is increased, and consequently a gap in the $I$-$V$ curve opens. Such
a gap is as large as the one in the occupation, which is roughly $A_N$.

We emphasize that, despite the simplicity of the 
function used to mimic the discontinuity, our model
restores the correct $I$-$V$ behavior with the expected
conductance gap, roughly equal to the $A_N$, thus repairing the
faulty LDA description.  The lack of eigenvalue
discontinuity causes a dramatic overestimation
of the metallicity for a molecular junction obtained within LDA.
This result generalizes arguments previously given using many-body
theory \cite{EWK04}, or, within DFT, only for weak bias
\cite{kieron}.

In contrast, the curves obtained for LDA and DISC in the strong coupling 
limit (figures \ref{Fig2}(a2), \ref{Fig2}(b2) and \ref{Fig2}(c2)) look rather similar.
This is because a considerable fraction of the level occupation and the current 
comes from the tail of the energy level DOS. In the large coupling limit,
$\Gamma\sim A_N$, both $n$ and the current $I$ are rather insensitive to 
$\epsilon(n)$, and we find that standard continuous functionals give rather accurate 
$I$-$V$ characteristics.

Having identified the lack of the derivative discontinuity in LDA as a major source of 
error in DFT-based transport calculations, we propose a corrective scheme
for the NEGF method. The key consideration is that in LDA (or GGA),
the linear behavior of the KS eigenvalues and the absence of derivative 
discontinuity in the total energy functional, is mainly due to the presence of 
self-interaction error (SIE), that is, the interaction of an electron with the 
exchange and correlation potential generated by its own charge \cite{PZ}.
This spurious interaction is responsible for a series of failures of DFT.
Most notably, negatively charged ions are unstable and in general 
$-\epsilon_\mathrm{HOMO}^\mathrm{KS}$ is not even close to the ionization potential. 
The elimination of the SIE improves considerably the agreement with experiments 
with respect to LDA and, more importantly in this context, makes the KS 
eigenvalues resemble more closely the true removal energies \cite{PZ}. 

\begin{figure}[ht]
\begin{center}
\includegraphics[width=6.5cm,clip=true]{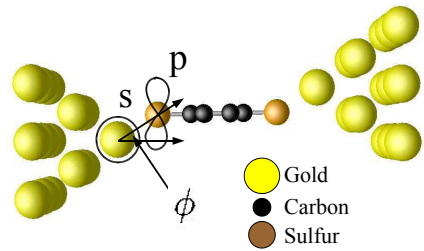}
\end{center}
\caption{\small{BDT molecule attached to (001) oriented Au surfaces.  
The angle $\phi$ between the BDT plane and the gold modulates the strength 
of the molecule/lead coupling.}}
\label{Fig3}
\end{figure}

Direct subtraction of SI in atoms is conceptually and practically simple \cite{PZ}.
However, the application of the method to extended systems is both cumbersome and 
computational demanding \cite{Tem}. A useful 
alternative is that first proposed by Vogel and then extended by Filippetti 
in which the SI is parametrized in terms of its atomic
counterparts and subtracted out (Pseudo-SIC, PSIC) \cite{pseudoSIC}. 
In the spirit of this method we have constructed an effective tight-binding model, 
and investigated the transport of a benzene-(1,4)-dithiolate (BDT) molecule 
sandwiched between two (001) oriented gold current/voltage probes 
(Fig.~\ref{Fig3}). This represents a benchmark, 
since LDA-DFT calculations \cite{BDT-DFT} and experiments \cite{BDT-EXP} 
are in stark disagreement. In particular most of 
the calculations fail to predict a conductance gap at zero bias. 

Here we adopt a minimal $\pi$ model where we consider only $p_z$ orbitals 
(orthogonal to the BDT plane) for both C and S atoms and $s$ orbitals for Au. 
H atoms are simply used for passivation and are not considered explicitly. 
The on-site energy of such $p_z$ orbitals are parameterized from their atomic
counterparts and coincide with the HOMO state of the free atom. This is computed for
different occupations with a standard self-consistent calculation using either 
LDA and SIC \cite{PZ}. The resulting $\epsilon(n)$ curves (not reported here) are similar to 
that of figure \ref{Fig1}(b). Our procedure neglects the crystal field, 
and assumes that the electron screening is weak. Although for a fully quantitative analysis 
such aspects must be considered, we do not expect that these details will change the main
features of our model. Finally, the hopping integrals are taken from the literature \cite{Har}.

The $I$-$V$ characteristics are then calculated using standard NEGF methodology with a tight-binding
version of our code {\it Smeagol} \cite{smeagol,TBSmeagol}.
In the simulation, we alter the strength of the coupling to the 
leads by varying the angle $\phi$ between the BDT plane and the apex of the Au pyramid 
(see figure \ref{Fig3}). The coupling is then $\gamma\sin\phi$ with
$\gamma$ the Au-S $sp\sigma$ hopping integral. The alignment of $E_\mathrm{F}$
of the leads with $\epsilon_\mathrm{HOMO}$ of the isolated molecule has been chosen in order to
reproduce that calculated by DFT-LDA, although variations of $\pm1$~eV around this value do
not cause any significant change in our results.

\begin{figure}[ht]
\begin{center}
\includegraphics[width=7.0cm,clip=true]{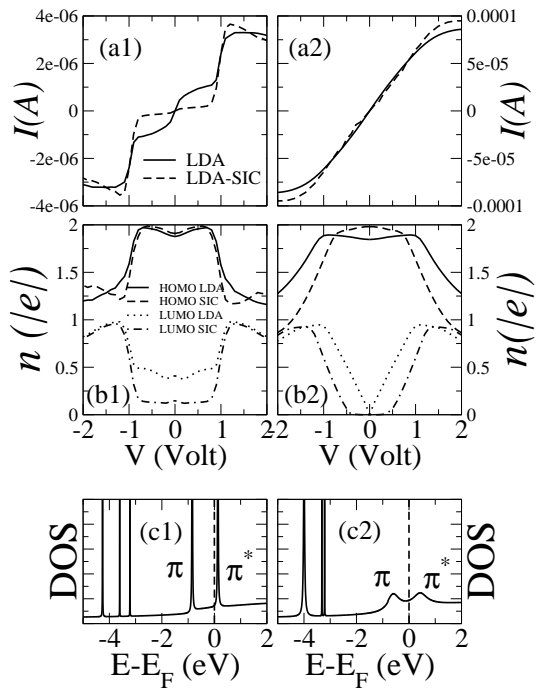}
\end{center}
\caption{\small{(a) $I$-$V$ characteristic, (b) occupation as a function of energy, and (c) 
DOS for the system BDT+Au leads. The cases of strong ($\phi=30^o$) and weak coupling 
($\phi=5^o$) are presented respectively in the right- and left-hand side panels.
We present only LDA DOS, since PSIC gives almost identical results. In panels (c) the 
vertical lines denotes the position of $E_\mathrm{F}$.}}
\label{Fig4}
\end{figure}

In Fig. \ref{Fig4}, we present our calculated $I$-$V$ curves, 
the occupation of the HOMO and LUMO state as a function of bias $V$,
and the DOS for both the weak ($\phi=5^o$) and strong ($\phi=30^o$) 
coupling regime.
For weak coupling, LDA and PSIC give dramatically different $I$-$V$ characteristics.
In particular, PSIC opens a conductance gap in the $I$-$V$ around zero bias. 
This is despite that fact that the LDA and PSIC DOS look almost 
identical. In both cases $E_\mathrm{F}$ pins the bottom of the S-derived empty 
$\pi^*$ orbital, which is the first state to get involved in the transport process. Once bias
is applied, such a LUMO state gets progressively more populated and follows the lead kept
at positive bias. The current is then roughly proportional to the state occupation, as seen
previously in the case of the simple model. The key point here is that, while in LDA the state 
charges linearly with bias, in PSIC it can follow the upper bias without charging significantly. 
Again the onset of charging will become important only when the state has moved upwards in energy
enough to match the derivative discontinuity. At this point the LUMO $\pi^*$
state starts to conduct (around $V$=1~Volt). In addition, for such biases, also the HOMO 
$\pi$ state appears in the bias window and contributes to the current.

In the strong coupling limit, the differences between LDA and PSIC are much less evident. 
In this case, both the $\pi$ and $\pi^*$ states are very broad (see figure \ref{Fig4}(b2)) 
providing contributions to the current, even at low bias. 

We emphasize here that a more rigorous approach would be to
do, for example, an exact exchange calculation \cite{KKP04} within NEGF.
This is within present computational capability, but would be
costly.  Our results demonstrate that such a calculation would
yield very different results from LDA or GGA calculations in the
weak coupling limit.

In conclusion, we have discussed the main characteristics of DFT-based NEGF 
methods. We have identified the lack of derivative discontinuity in continuous
density functional approximations
as a major source of error in calculating the $I$-$V$ characteristic of a
molecular junction. Our results demonstrate that {\em LDA and GGA are not
suitable for transport calculations}, at least when the coupling is weak.
We have further proposed a simple corrective scheme based on the
removal of the atomic self-interaction. This has the remarkable property of
re-introducing, albeit in an approximate way, the derivative discontinuity
of the potential, while adding moderate additional computational costs. These
KS eigenvalues are more closely related to the true removal energies, and 
therefore can be employed in a NEGF transport calculation. We have implemented such
a method in a simplified tight-bonding scheme and demonstrated that conductance gaps
at low bias can open for molecular junctions predicted metallic by LDA. 

We thank T.~Todorov and F.~Evers for useful discussions.
This work is funded by the Science Foundation of Ireland (SFI02/IN1/I175). 
KB was also supported by the US Department of Energy (DE-FG02-01ER45928).

{\small{

}}


\begin{thebibliography}{99}

\bibitem{gates1}
C.P.~Collier et al., Science {\bf 285}, 391 (1999); Y.~Huang et al., 
Science {\bf 294}, 1313 (2001), 

\bibitem{chem}
Y.~Cui, W.~Qingqiao, P.~Hongkun and C.M.~Lieber,
Science {\bf 293}, 1289 (2001);
F.~Patolsky et al.,
Proc. Natl. Acad. Sci. USA {\bf 101}, 14017 (2004).

\bibitem{negf}S. Datta, {\it Electronic Transport in Mesoscopic Systems}, (Cambridge University Press,
Cambridge, UK, 1995).

\bibitem{dft} 
H.~Hohenberg and W.~Kohn, 
Phys. Rev. {\bf 136}, B864 (1964)

\bibitem{ks} 
W.~Kohn and L.J.~Sham, 
Phys. Rev. {\bf 140}, A1133 (1965).

\bibitem{smeagol}A.~Reily Rocha, V.~M.~Garcia~Su\'arez, S.~W.~Bailey, C.~J.~Lambert, J.~Ferrer
and S.~Sanvito, Nature Material {\bf 4}, 335 (2005);
M.~Brandbyge et al., Phys. Rev. B {\bf 65}, 165401 (2002); J.~Taylor, H.~Guo, and 
J.~Wang, Phys. Rev. B {\bf 63}, 245407 (2001).

\bibitem{EWK04}
F. Evers, F. Weigend, M. Koentopp, Phys Rev. B {\bf 69}, 235411 (2004).
                                                                               
\bibitem{kieron}K. Burke, M. Keontopp and F. Evers, cond-mat/0502385.

\bibitem{SZVD05}
N. Sai, M. Zwolak, G. Vignale, and M. Di Ventra, Phys. Rev. Lett. {\bf 94}, 186810 (2005).

\bibitem{MB}M.H. Hettler, W. Wenzel, M.R. Wegewijs and H. Schoeller, Phys. Rev. Lett. {\bf 90},
076805 (2004); P. Delaney and J.C. Greer Phys. Rev. Lett. {\bf 93}, 036805 (2004); B. Muralidharan, A.W. Ghosh,
S.K. Pati and S. Datta, cond-mat/0505375

\bibitem{janak}J.F. Janak, Phys. Rev. B {\bf 18}, 7165 (1978).

\bibitem{PPLB}J.H. Perdew, R.G. Parr, M. Levy and J.L. Balduz Jr., Phys. Rev. Lett. {\bf 49}, 1691 (1982);
J.H. Perdew and M. Levy, Phys. Rev. Lett. {\bf 51}, 1884 (1983).

\bibitem{PL2}J.H. Perdew and M. Levy, Phys. Rev. B {\bf 56}, 16021 (1997).

\bibitem{Pb85}
J.P. Perdew, in  {\sl Density  Functional  Methods
in Physics}, edited by R.M. Dreizler and J. da Providencia  (Plenum, NY, 1985),
p. 265.
                                                                                
\bibitem{Alessio}A. Filippetti, Phys. Rev. A {\bf 57}, 914 (1998).

\bibitem{PZ} J.H. Perdew and A. Zunger, Phys. Rev. B {\bf 23}, 5048 (1981).

\bibitem{Tem} A. Svane and O. Gunnarsson, Phys. Rev. Lett. {\bf 65}, 1148 (1990); Z. Szotek, 
W.M. Temmerman and H. Winter, Phys. Rev. B {\bf 47}, 4029 (1993).
	
\bibitem{pseudoSIC} D. Vogel, P. Kr\"uger and J. Pollmann, Phys. Rev. B {\bf 54}, 5495 (1996);
A. Filippetti and N.A. Spaldin, Phys. Rev. B {\bf 67}, 125109 (2003)

\bibitem{BDT-DFT}M. Di Ventra, S.T. Pantelides and N.D. Lang, Phys. Rev. Lett. {\bf 84}, 979 (2000);
Y. Xue and M.A. Ratner, Phys. Rev. B {\bf 68}, 115406 (2003); Y. Xue and M.A. Ratner, Phys. Rev. 
B {\bf 68}, 115407 (2003)

\bibitem{BDT-EXP}M.A. Reed, C. Zhou, C.J. Muller, T.P. Burgin and J.M. Tour, Science {\bf 278}, 252 (1997)

\bibitem{Har}W.A. Harrison, {\it Electronic Structure and Properties of Solids}, (1980).

\bibitem{TBSmeagol} A. R. Rocha, S. Sanvito, Phys. Rev. B {\bf 70}, 094406 (2004).

\bibitem{KKP04}
S. Kummel, L. Kronik, and J.P. Perdew, Phys. Rev. Lett. {\bf 93}, 213002 (2004).                                                                                
                                                                                


\end{thebibliography}
\end{document}